\documentclass[aps, prfluids, reprint, onecolumn]{revtex4-2}

\usepackage{graphicx}
\usepackage{epstopdf, epsfig}
\usepackage{amsmath}
\usepackage[export]{adjustbox}

\usepackage{hyperref}
\hypersetup{
  colorlinks   = true, 
  urlcolor     = blue, 
  linkcolor    = blue, 
  citecolor   = green 
}

\usepackage[normalem]{ulem}
\usepackage{color}

\usepackage{tikz}
\definecolor{magenta_1}{rgb}{0.64,0.08,0.18}
\definecolor{yellow_1}{rgb}{0.93,0.69,0.13}

\definecolor{WCblue}{rgb}  {         0    0.4470    0.7410}
\definecolor{WCred}{rgb}   {    0.8500    0.3250    0.0980}
\definecolor{WCyellow}{rgb}{    0.9290    0.6940    0.1250}

\newcommand{\blackline} {\raisebox{2pt}{\tikz{\draw[-,black,solid,line width = 0.9pt](0,0) -- (5mm,0);}}}

\newcommand{\WCblueline} {\raisebox{2pt}{\tikz{\draw[-,WCblue,solid,line width = 0.9pt](0,0) -- (5mm,0);}}}
\newcommand{\WCredline} {\raisebox{2pt}{\tikz{\draw[-,WCred,solid,line width = 0.9pt](0,0) -- (5mm,0);}}}
\newcommand{\WCyellowline} {\raisebox{2pt}{\tikz{\draw[-,WCyellow,solid,line width = 0.9pt](0,0) -- (5mm,0);}}}

\begin{document}

\title{Widest scales in turbulent channels}

\author{R. Pozuelo}
\affiliation{FLOW, Engineering Mechanics, KTH Royal Institute of Technology, SE-100 44 Stockholm, Sweden}
\author{A. Cavalieri}
\affiliation{Divis\~{a}o de Engenharia Aeron\'autica, Instituto Tecnol\'ogico de Aeron\'autica (ITA), 12228-900, S\~{a}o Jos\'e dos Campos, SP, Brazil}
\author{P. Schlatter}
\affiliation{Institute of Fluid Mechanics (LSTM), Friedrich-Alexander-Universit\"at (FAU) Erlangen-N\"urnberg, Germany}
\author{R. Vinuesa}
\affiliation{ FLOW, Engineering Mechanics, KTH Royal Institute of Technology, SE-100 44 Stockholm, Sweden}

\date{\today}

\begin{abstract}
	The widest spanwise scales in turbulent channel flows are studied through the use of three periodic channel-flow simulations at friction Reynolds number $\mathrm{Re}_{\tau}=550$. The length and height of the channels are the same in all cases ($L_x/h=8\pi$ and $L_y/h=2$ respectively), while the width is progressively doubled: $L_z/h = \{4\pi, 8\pi, 16\pi\}$.
    The effects of increasing the domain can not be determined with statistical significance in our simulations, since the difference in the statistics between the simulations is of the same order as the errors of convergence.
    A channel flow similar to the smaller one (\textit{J.~Fluid Mech.}, vol. 500, 2004, pp. 135--144), which was averaged over a very long time, was used as a reference.
    The one-dimensional spanwise spectrum of the streamwise velocity is computed with the aim of assessing the domain-size effect on the widest scales. Our results indicate that $90\%$ of the total streamwise energetic fluctuations is recovered without a significant influence of the size of the domain. The remaining $10\%$ of the energy reflects that the widest scales in the outer layer are the ones most significantly affected by the spanwise length of the domain. 
    The power-spectral density for $k_z = 0$ remains constant even if the size of the domain in the spanwise direction is increased up to 4 times the standard spanwise length, indicating that wide, spanwise coherent structures are not an artifact of domain truncation.
\end{abstract}

\maketitle

\section{Introduction}

In numerical simulations of turbulent flows, the size of the domain imposes a limit to the size of the scales that can be simulated in the domain, and the resolution is responsible for properly resolving the various simulated scales.
In the early days of direct numerical simulations (DNSs) of turbulent channel flows, \citet*{kim87} showed their concern for the smallest scales that could not be captured by the resolution they were using, and the domain size was justified by the use of two-point correlations.
The use of the one-dimensional energy spectra in the streamwise and spanwise directions was then used to justify the grid resolution, since the spectral energy for high wavenumbers decreased by orders of magnitude.
In this early simulation, the values of the energy spectra exhibited large energy at low wavenumbers (large scales).

Later, \citet*{jimenez1998largest} focused on the largest scales present in turbulence, where the criterion for a domain that could fit the largest scales relied on the function $k\phi$ (which is the power-spectral density $\phi$ premultiplied by the wavenumber $k$) decaying for large and small wavenumber. 
Such function decays, by construction, as $k$ goes towards zero.
The largest scales indeed contain a low spectral energy per unit of volume, but their big size can make their total energetic contribution to be a substantial fraction of the turbulence statistics. Their long lifetimes can also accumulate over time and they have important implications on the flow from an acoustic point of view \cite{Sano_2019}.
More studies on the effect of the size of the domain were conducted by \citet*{jimenez_moin_1991}, who defined the minimal flow unit. In those simulations the aim was to assess small domains that can capture ``healthy'' turbulence, implying that the turbulence statistics for a certain region close to the wall are correct. In \citet*{jimenez_moin_1991} the focus was on the minimal dimensions for the viscous and buffer layers, while in \citet*{Flores_Jimenez_2010} the study was extended to the logarithmic region. Those studies assessed the effects of small channels with different aspect ratios and those results were compared with highly-resolved turbulent channel flows at a similar friction Reynolds numbers, {\it i.e.} $\mathrm{Re}_{\tau}=180$ from \citet*{kim87} and $\mathrm{Re}_{\tau}=2000$ from \citet*{Hoyas_PoF2006}.
The smallest channels exhibit lower values of the streamwise root-mean-square (rms) fluctuations $u_{\mathrm{rms}}$ and larger values for the wall-normal component $v_{\mathrm{rms}}$.

Focusing on the largest scales in turbulent channel flows, where the premultiplied streamwise spectral density did not exhibit closed contours for the standard domain sizes being simulated, \citet*{Adrian_chan_2014} performed a simulation of a very large channel at $\mathrm{Re}_{\tau}=550$ with streamwise and spanwise dimensions $(L_x, L_z) = (60\pi h , 6\pi h)$, where $h$ is the half-channel height.
They showed that their one-point statistics were barely affected by increasing the width of the domain beyond $\pi h$ and their length beyond $2\pi h$ for $\mathrm{Re}_{\tau}$ up to 2000. The effects observed in the premultiplied two-dimensional spectra at the center of the channel showed that shorter domains exhibit trends similar to those for larger domains.
They were able to close the contour corresponding to $10\%$ of the maximum, which they report contains $80\%$ of the streamwise turbulent kinetic energy.
Note that they did observe some accumulation of energy in the probability density functions of the coherent structures studied in \citet*{Lozano_coherent_2012}. For shorter domains, the largest coherent scales exhibited larger energy levels in comparison with the larger domains. They argued that the largest resolved scales and those that were too large to be simulated in the domain, which was seen as infinite in size due to the periodicity, did not alter the dynamics of the rest of the simulated scales. 
Motivated by the unexplored spanwise large scales and the studies that related these scales with the noise generation at the trailing edge of a NACA profile, 
Abreu {\it et al.}~\cite{Abreu_JFM_2021} 
studied these spanwise coherent structures. They observed that the power-spectral density in zero-pressure-gradient (ZPG) turbulent-boundary-layer (TBL) simulations collapsed for spanwise wavenumbers $k_z \rightarrow 0$, when the domain was increased in the spanwise periodic direction. The use of SPOD showed that the same hydrodynamic spanwise structures appeared in NACA profiles, and argued that these waves are not an artifact of the periodic conditions or the constrains in the spanwise direction of the simulated domain.

The purpose of the present work is to quantify the effects of the widest structures in turbulent channel flows. We start from a highly-resolved case similar to the one used in Del \'Alamo {\it et al.}~\cite{delAlamo_jfm_2004} for $\mathrm{Re}_{\tau}=550$, and simulate two more channels at the same $\mathrm{Re}_{\tau}$, with the same streamwise length and similar sampling conditions, with the only difference being the width of the domain, which is doubled from one channel to another.

\section{Numerical setup} \label{sec:NumSetUp}

The parameters of the simulations used in this work are collected in Table~\ref{tab:param_sch}, where the physical resolution of the domain is maintained in all dimensions, with the only difference being the spanwise size of the domain $L_z$ which is doubled subsequently for channels C2 and C3.
The resolution is reported in viscous units, which are denoted by the superscript `+', where the fluid kinematic viscosity $\nu$ and the friction velocity $u_{\tau}=\sqrt{\tau_w / \rho}$ ($\tau_w$ is the wall-shear stress and $\rho$ the fluid density) are employed.
The evolution equations for the incompressible flow are integrated using the formulation based on the wall-normal vorticity and the Laplacian of the wall-normal velocity for the current simulations performed with SIMSON \citep{simson_techrep, li2009simson2d}.
The domain is discretized using Fourier modes in the periodic streamwise $x$ and spanwise $z$ directions. Chebychev polynomials are used in the wall-normal direction $y$. 
For the channel case ``torr'', the employed time integration was a third-order semi-implicit Runge--Kutta scheme, while the one used for channels C1, C2 and C3 was a four-stage Runge--Kutta scheme for the advective
term, and a second-order Crank--Nicolson scheme for the viscous terms for all the simulations.
The time-step of the simulation, as well as the sampling times for the statistics and the velocity fields used to calculate the power spectrum of the velocity fluctuations, are shown in Tab.~\ref{tab:param_time}.
Note that the smaller channels torr and C1 have the same dimensions and resolution, and the time span of the torr case is slightly larger than C1.
We expect to obtain similar one-point statistics in all these simulations. Since we will analyze the difference in these quantities due to different size of the domains, we have also included the simulation ``torr'' by \citet*{Alamo_spec_2003_PoF} to illustrate how big can the differences be for the same domain size. 
\begin{table*}
    \begin{center}
    \def~{\hphantom{0}}
    \begin{tabular}{ l r  r  r  r  r  r  r  r  r  r   r }
    Case  & $\mathrm{Re}_{\tau}$ & $L_x/h$ & $L_y/h$ & $L_z/h$ & $m_{x}$ & $m_{y}$ & $m_{z}$ & $\Delta x^{+}$ & $\Delta y_{\rm wall}^+$ & $\Delta z^{+}$ & Colour        \\[3pt]
    \hline
    torr  &     547     & 8$\pi$  &    2    & 4$\pi$  &   1536  &   257   &   1536  &      8.9       &            0.041        &       4.5      & \blackline    \\
    C1    &     545     & 8$\pi$  &    2    & 4$\pi$  &   1536  &   257   &   1536  &      8.92      &            0.041        &       4.46     & \WCblueline   \\
    C2    &     545     & 8$\pi$  &    2    & 8$\pi$  &   1536  &   257   &   3072  &      8.92      &            0.041        &       4.46     & \WCredline    \\
    C3    &     545     & 8$\pi$  &    2    & 16$\pi$ &   1536  &   257   &   6144  &      8.92      &            0.041        &       4.46     & \WCyellowline \\

    \end{tabular}
    \caption{Parameters of the simulations used in this paper. The reference length $h$ corresponds to the channel half-height. The box size is $(L_x/h, L_y/h , L_z/h)$  and $(m_x, m_y, m_z)$ are the number of collocation points including the $3/2$ factor for de-aliasing in the Fourier directions ($x$ and $z$). The spatial resolution in viscous units of the streamwise and spanwise directions $\Delta x^{+}, \Delta z^{+}$ has been calculated using $(m_x, m_z)$. The largest grid spacing in the wall-normal direction is located at the center of the channel, and it is equal to $\Delta y_{\rm max}/h=0.0123$ or $\Delta y_{\rm max}^+ = 6.69$. }
    \label{tab:param_sch}
    \end{center}
\end{table*}

\begin{table*}
    \begin{center}
    \def~{\hphantom{0}}
    \begin{tabular}{l  c  c  c  c  c  c  c  c  c  c}
    Name & $\Delta t_{\rm sim}$ & $\Delta t_{\rm stats}$   & $\Delta t_{\rm spec}$   & $\Delta t_{\rm sim} ^+$ & $\Delta t_{\rm stats} ^+$   & $\Delta t_{\rm spec} ^+$    & Samples spec & $\Delta T$  & $\Delta T / (h/u_{\tau})$ & $\Delta T U_b/ L_x$   \\
    \hline
    torr &         --       &           --         &          --         &             --      &             --          &             --          &    --        &  --         &                 12.3     &                 10    \\
    C1   &         0.004    &         0.064        &         0.64        &           0.079     &            1.26         &            12.6         &   469        & 300         &                 10.9     &                 7.96  \\
    C2   &         0.004    &         0.064        &         0.64        &           0.079     &            1.26         &            12.6         &   469        & 300         &                 10.9     &                 7.96  \\
    C3   &         0.004    &         0.064        &         0.64        &           0.079     &            1.26         &            12.6         &   469        & 300         &                 10.9     &                 7.96  \\
    \end{tabular}
        \caption{Time steps for the simulation (sim), statistics sampling (stats) and spectrum sampling (spec) of the new simulations, where $\Delta T$ is the total simulated time. Viscous units are denoted with `+', $h$ is the channel half-height, $u_{\tau}$ is the friction and $U_b$ is the bulk velocity.}
        \label{tab:param_time}
    \end{center}
\end{table*}
\section{Turbulence statistics} \label{sec:Stats}
To assess the effects of the size of the domain we first evaluate its effect on the overall turbulence statistics.
The turbulence statistics for the channel flows are obtained through the averaging over time and the homogeneous wall-parallel directions $x$ and $z$ as well as in the wall-normal direction
using the property of symmetry or anti-symmetry of the measured quantity around the center of the channel.
Note that torr has been averaged over a longer time period than the rest of the simulations, and C1, C2, C3 have been averaged over the same time span.
Since C2 has double the size of C1, it will contain double of samples than C1 for the same averaged time.
The same can be applied to channel C3, which contains a larger number of samples than the other channels.
In Fig.~\ref{fig:WC_stats_U_torr} we examine the effects of the width of the domain on the streamwise mean velocity.
Fig.~\ref{fig:WC_stats_U_torr}(a) shows the inner-scaled streamwise-velocity profile, and to illustrate the differences between the various simulations we show 
in Fig.~\ref{fig:WC_stats_U_torr}(b) the absolute difference of each cases with respect to channel C3. 
The relative deviation of the mean streamwise velocity is represented in Fig.~\ref{fig:WC_stats_U_torr}(c), showing a value of the order of $0.1\%$.

\newcommand\sizeImg{0.32}
\graphicspath{{./Stats/torr/}}
\begin{figure}
\centering
\includegraphics[width=\sizeImg \textwidth]{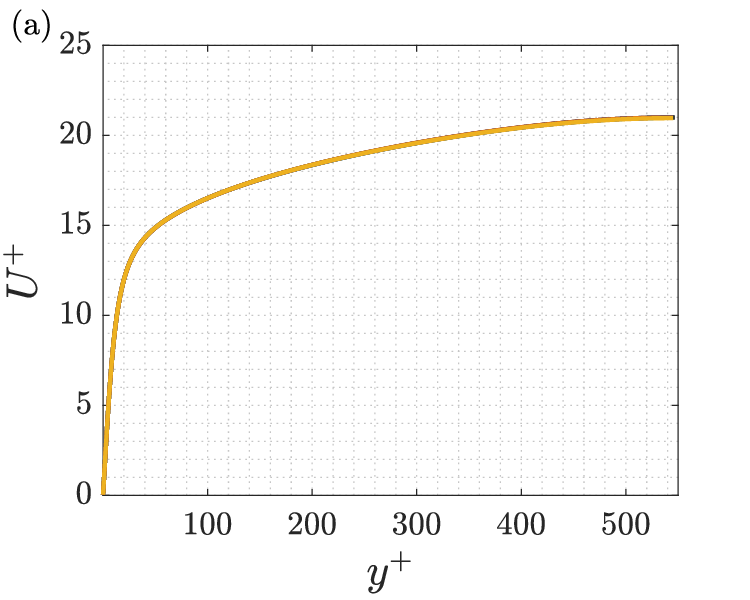} 
\includegraphics[width=\sizeImg \textwidth]{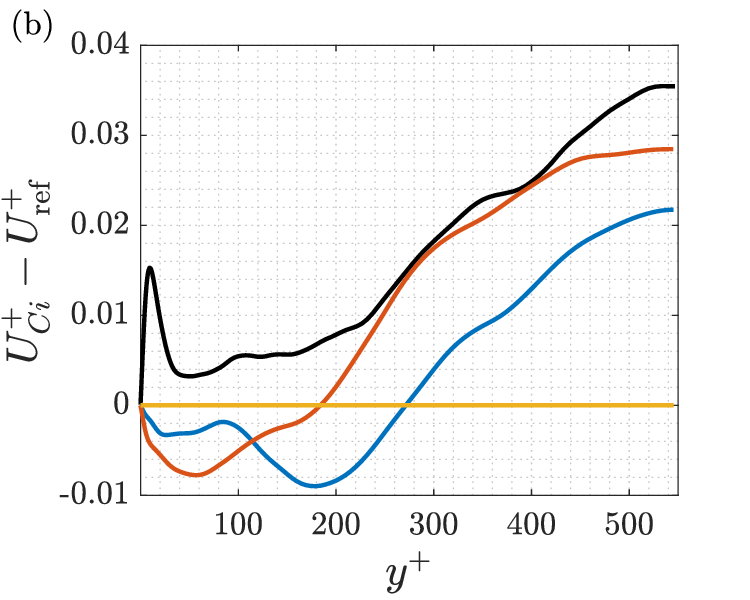} 
\includegraphics[width=\sizeImg \textwidth]{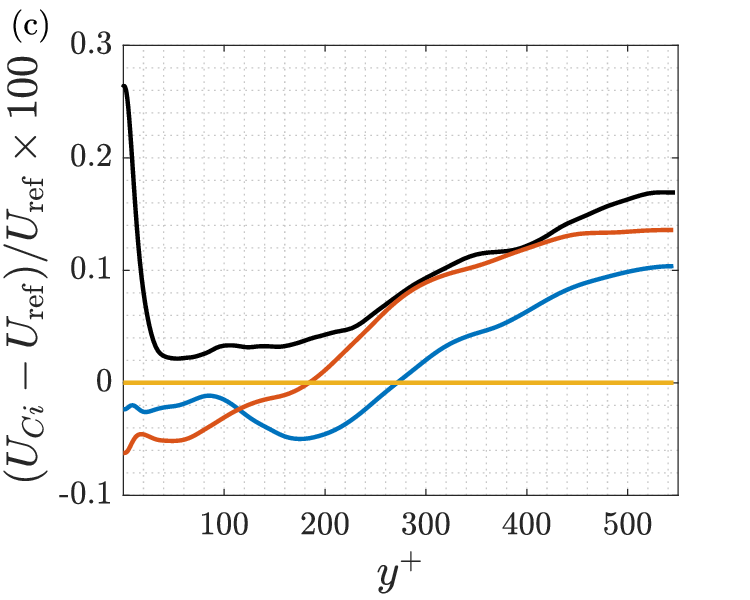}
\caption{ \label{fig:WC_stats_U_torr} (a) Mean streamwise velocity $U$ for channels C1 (blue), C2 (red), C3 (yellow) and torr (black) scaled in viscous units.
(b) Absolute deviation of $U$ with respect to the values of channel C3. (c) Relative deviation of $U$ with respect to the values of channel C3.
The data for the simulation torr was extracted from \citet*{Alamo_spec_2003_PoF}.
}
\end{figure}
\begin{figure}
\centering
\includegraphics[width=\sizeImg \textwidth]{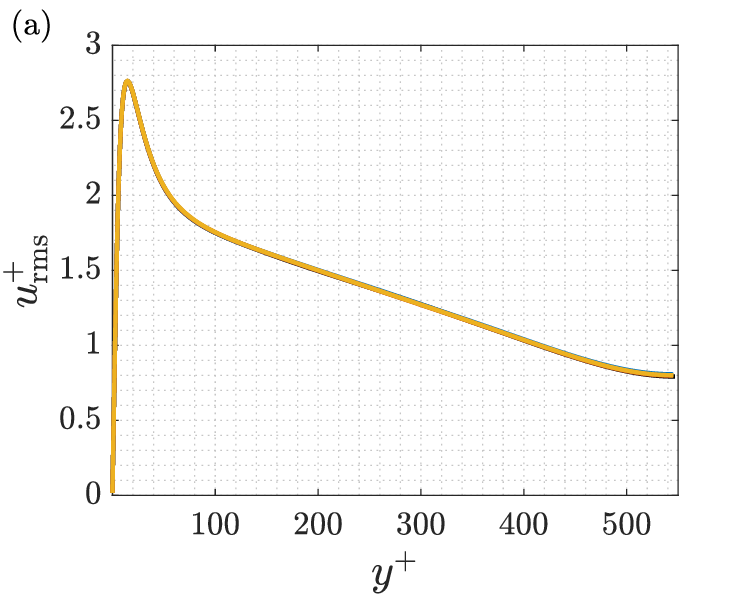}
\includegraphics[width=\sizeImg \textwidth]{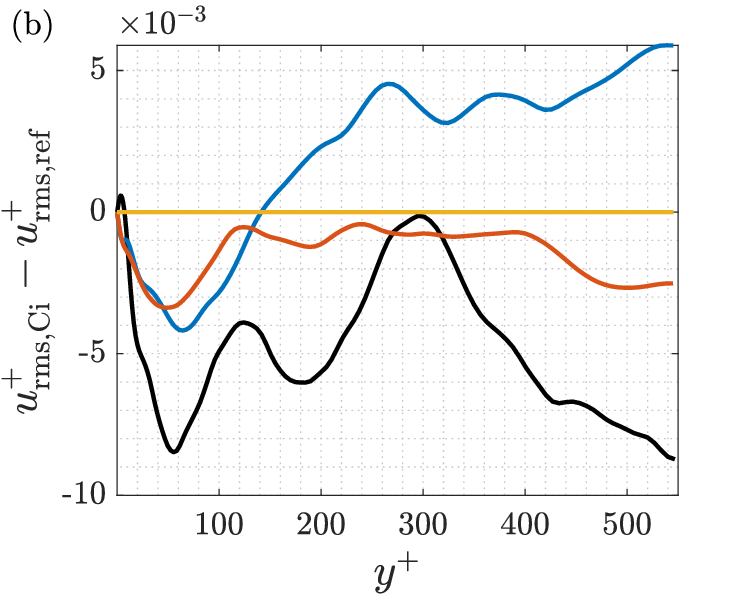}
\includegraphics[width=\sizeImg \textwidth]{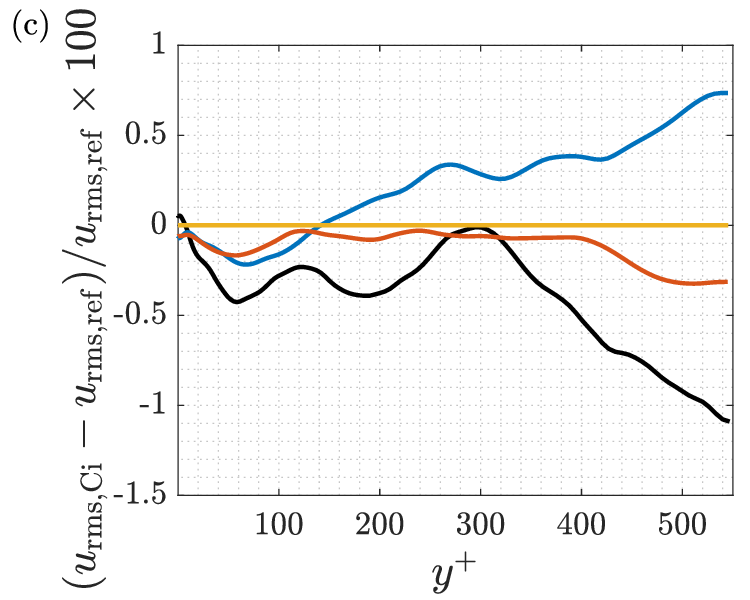}
\caption{ \label{fig:WC_stats_urms_torr} (a) Streamwise fluctuation profile $u^+_{\mathrm{rms}}$ for channels C1 (blue), C2 (red), C3 (yellow) and torr (black) scaled in viscous units.
(b) Absolute deviation of $u_{\mathrm{rms}}$ with respect to the values of channel C3. (c) Relative deviation of $u_{\mathrm{rms}}$ with respect to the values of channel C3.
The data for the simulation torr was extracted from \citet*{Alamo_spec_2003_PoF}.
}
\end{figure}

\begin{figure}
\centering
\includegraphics[width=\sizeImg \textwidth]{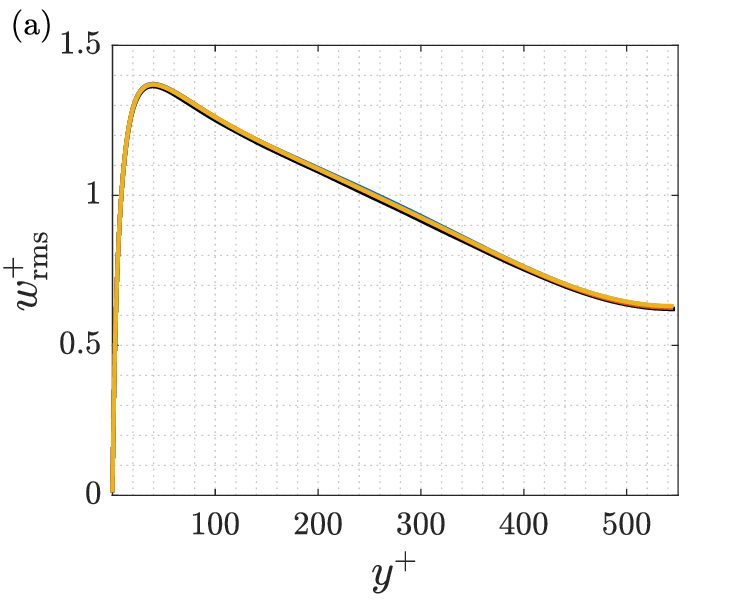}
\includegraphics[width=\sizeImg \textwidth]{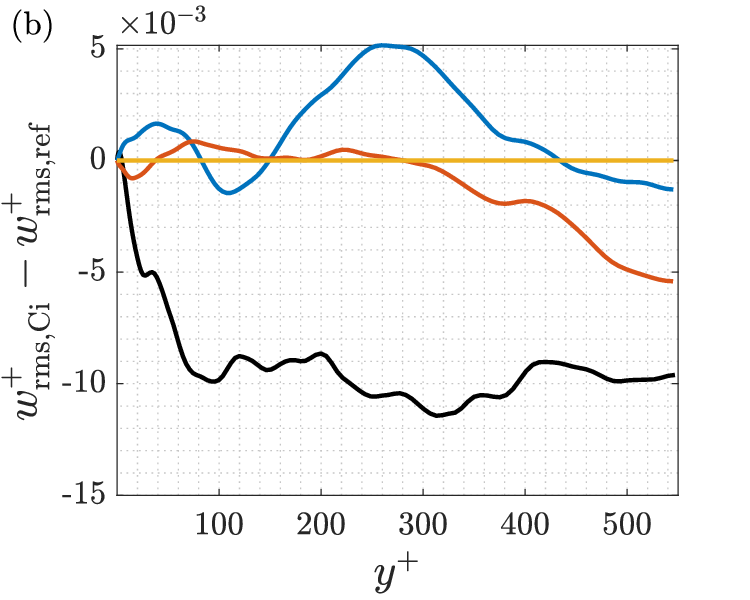}
\includegraphics[width=\sizeImg \textwidth]{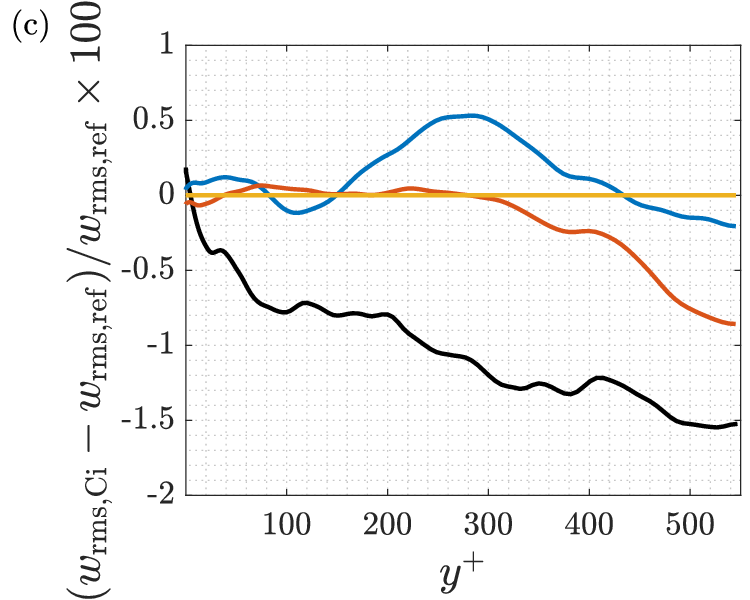}
\caption{ \label{fig:WC_stats_wrms_torr} (a) Spanwise fluctuation profile $w^+_{\mathrm{rms}}$ for channels C1 (blue), C2 (red), C3 (yellow) and torr (black) scaled in viscous units.
(b) Absolute deviation of $w_{\mathrm{rms}}$ with respect to the values of channel C3. (c) Relative deviation of $w_{\mathrm{rms}}$ with respect to the values of channel C3.
The data for the simulation torr was extracted from \citet*{Alamo_spec_2003_PoF}.
}
\end{figure}
As discussed by \citet*{Townsend_1976}, the presence of the wall limits the wall-normal perturbations, and the effects on the Reynolds stress components such as $v^2$ or $uv$ should be mainly observed in the outer region. 
In \citet*{Alamo_spec_2003_PoF} they report that the longest fluctuating scales appear for the streamwise velocity at $y=0.5h$ while the widest ones appearing for the spanwise velocity are at the center of the channel. For this reason, we will analyze the effect of the width of the domain in these components.
In Figs.~\ref{fig:WC_stats_urms_torr}~and~\ref{fig:WC_stats_wrms_torr} we show the rms of the streamwise and spanwise velocity fluctuations, together with the deviation and relative deviation with respect to C3.
The relative deviations, in this case, are of the order of $1\%$, with a larger difference in the outer region.
Based on these results and the order of magnitude of the errors, from a point of view of the turbulence statistics, the channels wider than C1 for $\mathrm{Re}_{\tau}=550$ do not introduce any significant difference.

\section{Spectrum of the widest turbulent scales} \label{sec:Spec}

To assess whether wider domains constrain the widest scales, we use the one-dimensional spectrum of the fluctuation velocities in the spanwise direction.
In the turbulence statistics, the averaging in the spanwise direction implies that C3 has in principle better convergence than channels C1 and C2. However, the average of the spanwise spectrum is performed only in time, the streamwise direction and the symmetry around the center of the channel. 
Since the time sampling and the streamwise size of the domain are the same for simulations C1, C2 and C3, the energy contained in the different spanwise scales are averaged over the same number of samples, and the convergence should be similar. 

The power spectrum is defined as $\mathrm{PS}=\langle \hat{u}(k_z, y)\hat{u}^*(k_z, y) \rangle_{(x,t,Sy)}$, where $\hat{u}$ indicates the Fourier coefficient of $u$ and $\hat{u}^*$ the complex conjugate, while the power-spectral density is $\phi(k_z, y) = \mathrm{PS}/\mathrm{d}k_z$.
Here $\langle \cdot \rangle$ denotes averaging over the quantities shown in the subscript 
and `Sy' indicates the symmetry applied in the wall-normal direction.
The widest scale captured by Fourier analysis corresponds to $k_z=(2\pi)/L_z$ or $\lambda_z = L_z$, and the energy contained in $k_z=0$ corresponds to the energy of spanwise constant fluctuations. 
\graphicspath{{./PS/}}

\begin{figure}
\centering
\begin{tabular}{@{}c@{}}
{\includegraphics[width=0.54\linewidth]{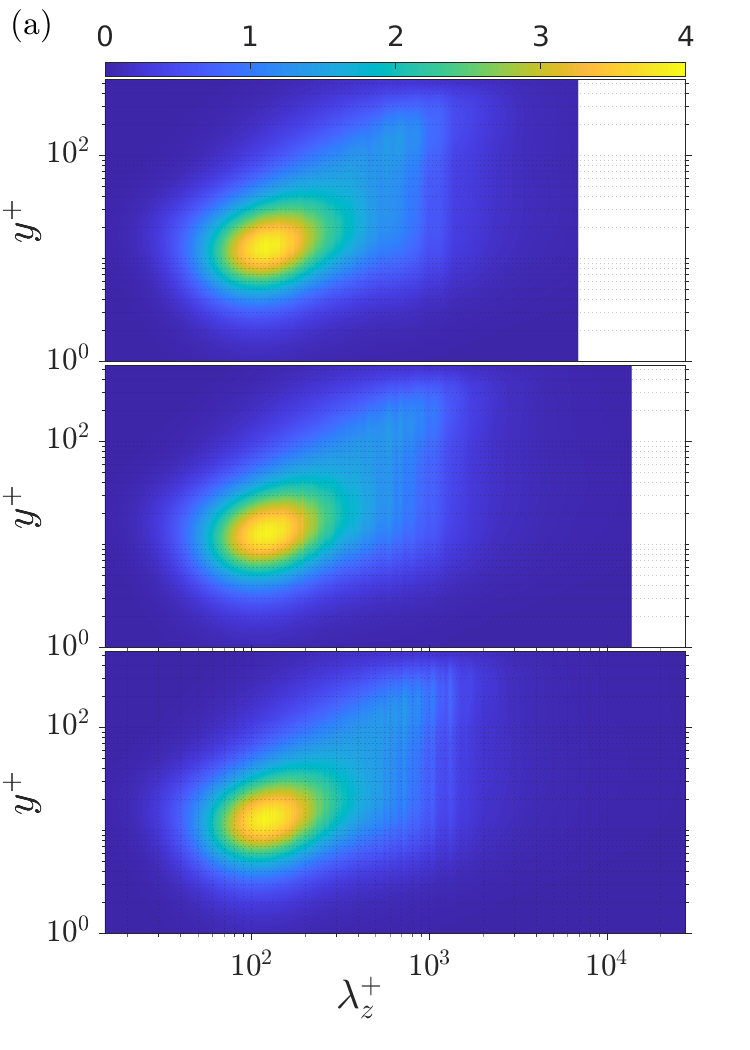}} 
\end{tabular}
\begin{tabular}{@{}c@{}}
{\includegraphics[width=0.45\linewidth]{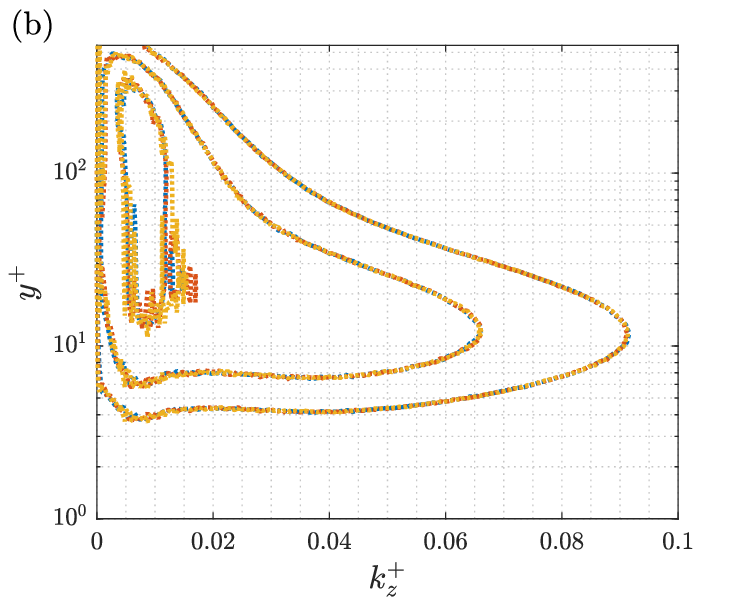}}
\\
{\includegraphics[width=0.45\linewidth]{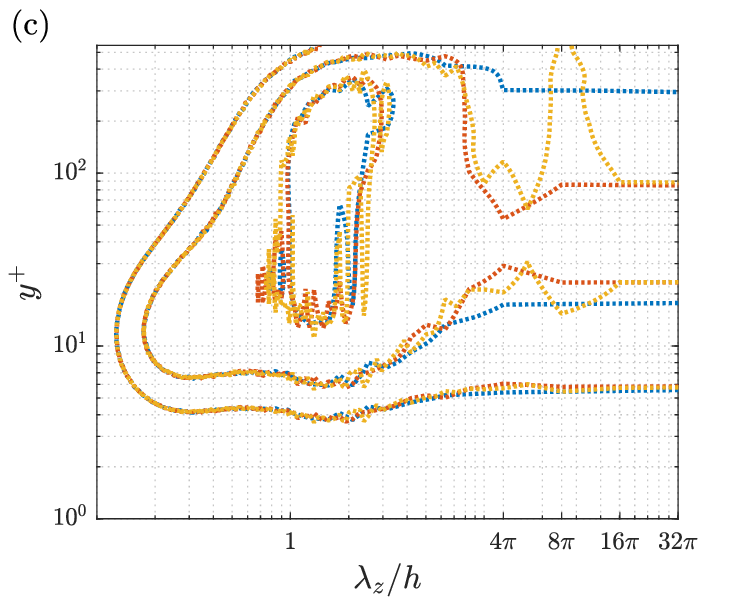}} 
\end{tabular}
\caption{(a) Inner-scaled premultiplied power-spectral density $k_z\phi_{u  u }/u_{\tau}^2$ as a function of the spanwise wavelength $\lambda_z^+$ for C1 (top), C2 (middle), C3 (bottom). (b) Inner-scaled power spectral density $\phi_{u  u }/u_{\tau}^2$ as a function of the spanwise wavenumber $k_z$; contours taken at $\phi_{u  u }^+ = [0.05, 0.1, 0.2]$, which correspond to $15\%$, $31\%$ and $62\%$ of the maximum. (c) Same as (b) but represented as a function of $\lambda_z$. Colors shown in Table~\ref{tab:param_sch}.}
\label{fig:kzPSD_PSD}
\end{figure}

In the spectral analysis of turbulence, the spectrum is usually shown premultiplied by the wavenumber. In Fig.~\ref{fig:kzPSD_PSD}(a) we show the premultiplied spectrum of the streamwise velocity for the three channel flows C1 (top), C2 (middle) and C3 (bottom), where it is possible to observe similar characteristics and values of the energy, with the only differences being the size of the scales that are captured and the spectral resolution being finer in the wider channels.
In Fig.~\ref{fig:kzPSD_PSD}(b) we represent the contours of $\phi_{u  u }$, without premultiplication by the wavenumber, as a function of the wavenumber, and the levels $31\%$ and $15\%$ reach the wavenumber $k_z = 0$, a fact that is manifested in Fig.~\ref{fig:kzPSD_PSD}(c) as contours which are not closed since $k_z=0$ corresponds to $\lambda_z=\infty$, or scales wider than the domain.

In the spectra discussed above, the integral of $\phi$ over all the wavenumbers (including the energy contained in the wavenumber $k_z = 0$), equals the corresponding Reynolds-stress component.
The premultiplication by $k_z$ automatically cancels that energy, making it inappropriate to use premultiplied spectra for integral analysis. 
The question is whether it is possible to have a domain $L_z$ where the largest Fourier scales $\lambda_z=\{L_z, L_z/2, ... \}$ do not contain spectral energy, thus leading to closed contours.
It is important to analyze the evolution of the spectral density of energy contained in $k_z=0$ with $L_z$, and even more relevant from the physical point of view is how much is the density of energy contained in the first wavenumbers or the largest scales $\lambda_z=\{L_z, L_z/2, L_z/3,...\}$, etc.

When comparing the spectra in channels of different widths, the wider the domain, the finer the wavenumber resolution. The smallest channels torr and C1 share a width of $L_{z} = 4\pi h$, which we will use as a parameter $L_{z,0} = L_{z,C1} = 4 \pi h$.
In channels torr and C1, the first wavelengths are $k_z = (2\pi/L_{z,0}) \{0, 1, 2, i, ... \}$, while for channel C2 they are $k_z = (2\pi/L_{z,0}) \{0, 1/2, 1, 3/2, 2, i/2, ... \}$ and for C3: $k_z = (2\pi/L_{z,0}) \{0, 1/4, 1/2, 3/4, 1, i/4, ... \}$. 
We see how channels C1 and torr have the same wavenumber resolution, while C2 and C3 have respectively 2 and 4 times the wavenumber resolution of C1.
We can compare the spectral density of energy obtained for scales of a similar width (same wavenumber/wavelength) for all the channels, and it will give an idea of how the spectral energy density changes due to numerical and time-average factors (comparing C1 and torr), or due to the increasing size of the domain, comparing C1 with C2 and C3.
\begin{figure}
    \centering
    \includegraphics[width=0.46\textwidth]{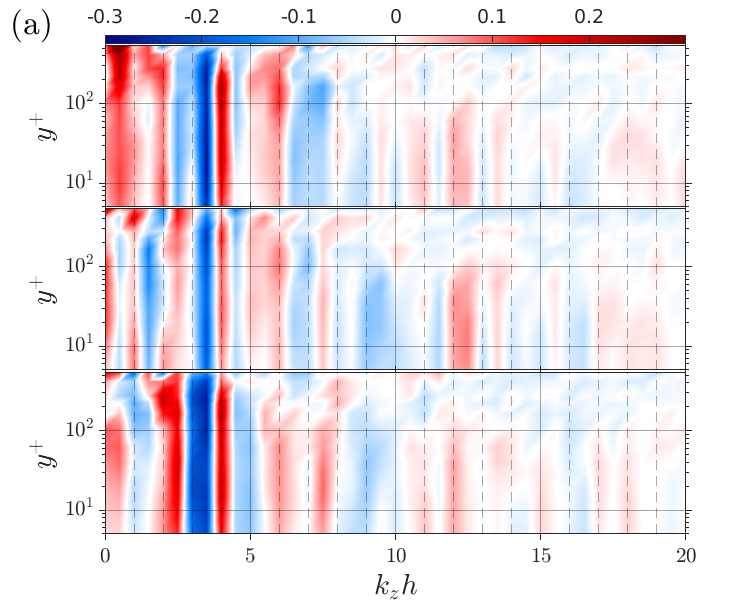}
    \includegraphics[width=0.46\textwidth]{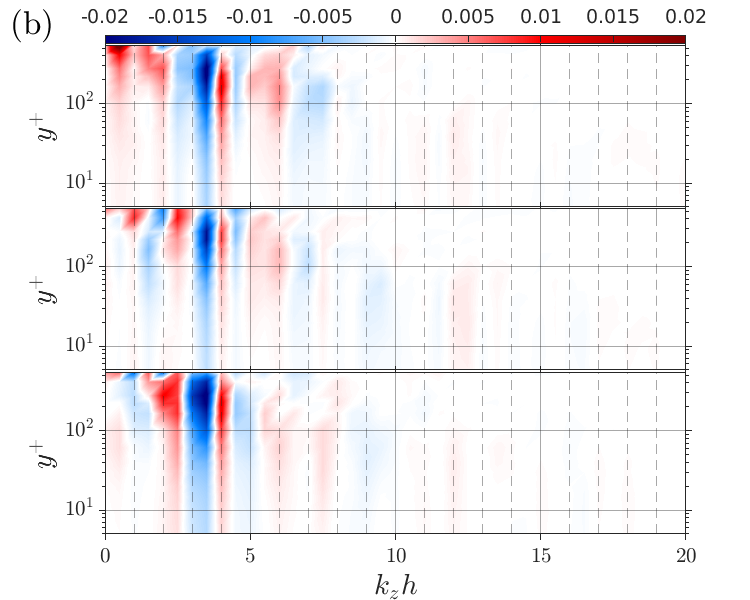}
    \caption{(a) Relative differences in power-spectral density with respect to that of torr channel $(\phi_{Ci}^+ - \phi_{\rm torr}^+)/\phi_{\rm torr}^+$ at the same wavenumbers and $y^+$. (b) Differences in power-spectral energy with respect to that of torr channel assuming the same bandwidth $\mathrm{d}k_{z, \mathrm{torr}}^+=\mathrm{d}k_{z, \mathrm{torr}}l_{\tau}$ and relative to the total energy as: $(\phi_{Ci}^+ - \phi_{\rm torr}^+)\mathrm{d}k_{z, \mathrm{torr}}^+/\int \phi_{\rm torr}^+ \mathrm{d}k_{z, \mathrm{torr}}^+$. The top row compares C1 with torr, the middle row is for C2 and the row at the bottom compares C3.}
    \label{fig:spec_diff_torr}
\end{figure} 

\begin{figure}
    \centering
    \includegraphics[width=0.46\textwidth]{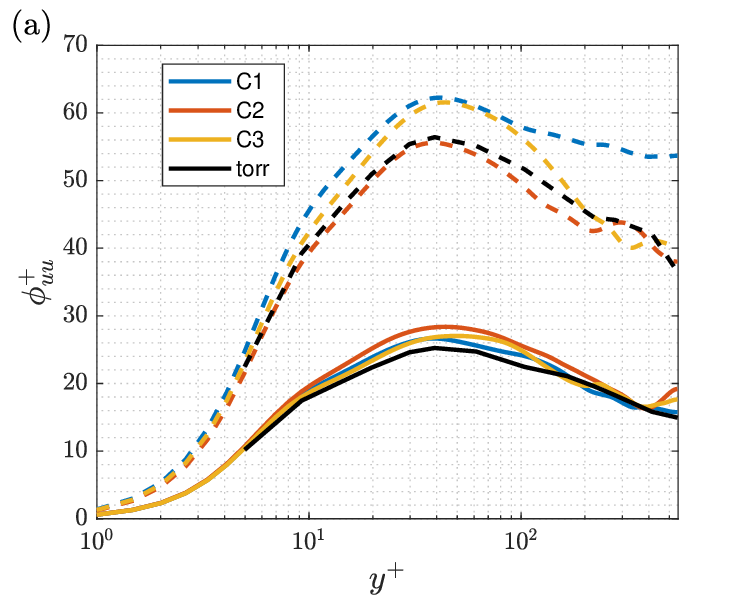}
    \includegraphics[width=0.46\textwidth]{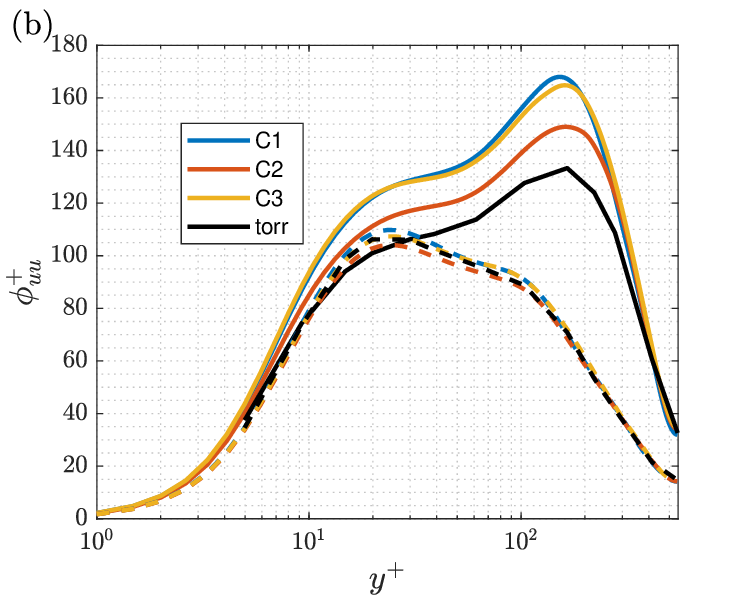}
    \caption{Power-spectral energy density $\phi_{uu}^+ $ for the different channels at wavenumbers (a) $k_z  h = 0$ (solid lines) and at $k_z  h = 0.5$ (dashed lines) for the different channels. (b) Power-spectral energy density at wavenumbers $k_z  h = 4$ (solid lines) and at $k_z  h = 8$ (dashed lines).}
    \label{fig:spec_kz_0_to_8}
\end{figure}

In Fig.~\ref{fig:spec_diff_torr}(a) we show the relative difference of $(\phi^+_{Ci} - \phi^+_{\rm ref}) / \phi^+_{Ci} $ where the reference $\phi^+_{\rm ref}$ has been chosen as $\phi^+_{\rm torr}$. Here, $\phi^+ = (\mathrm{PS}/u_\tau^2)(\mathrm{d}k_{z} l_{\tau})$ such that $\int \phi^+ \mathrm{d}k_{z}^+ = \overline{uu}^+$. 
The largest relative difference is of the order of $40\%$ and is located at $y=h$ for wavenumbers $k_z h = 0.5$ between torr and C1, but also between C1 and C2. The largest differences $\approx 30\%$ are located above $y^+=200$ for wavenumbers $k_{z} h <5$. Although these relative differences are large for the largest scales, mostly due to an insufficient amount of time to converge the spectra of these scales, it is also interesting to see how large are these differences compared to the total energy. In Fig.~\ref{fig:spec_diff_torr}(b) we assume a common wavenumber bandwidth $\mathrm{d}k_{z}^+ = \mathrm{d}k_{z, \mathrm{torr}}^+ = \mathrm{d}k_{z, C1}^+$ and we compare the differences in energy at similar wavenumbers with the total energy $\overline{uu}^+$. In this figure, it is easier to see how the largest differences are concentrated in the largest scales closer to the center of the channel, and they can amount up to $2\%$ of the total Reynolds stress.
Notice how the differences alternate in sign for different wavenumbers; this is also observed if the reference is taken as the spectral density for C1, C2 or C3. This result shows that the energy of scales that cannot be captured in narrower domains, is not accumulated in their largest scales. 
The same method has been used to compare channels C1 and C2 with C3 as a reference at matching wavenumbers/wavelengths. A similar pattern has been observed where the differences in power-spectral density alternate sign for different Fourier modes. 
The differences between C1 and C3 do not always share the same sign as the differences between C2 and C3.
For wider scales, we see larger differences that extend all across the channel height, and for shorter scales, the difference is smaller in magnitude and 
spans shorter regions in the buffer and logarithmic layers.
An example of $\phi^+$ is shown in Fig.~\ref{fig:spec_kz_0_to_8} where it can be seen that for $k_z = 0$ the level of $\phi^+$ is similar for all the simulations and for $k_z  h = 0.5$ the lines for C1 and the wider C3 are similar.
This indicates that the low $k_z$ content is not an artifact of the finite domain size, as the various domains considered here all lead to the same power-spectral density for the lowest spanwise wavenumbers. The largest Fourier scales always display non-zero power spectral density, even for wide channels.
For larger wavenumbers such as $k_z  h = 8$ the differences are smaller between the different channels, and their relative error decreases exponentially with the wavenumbers.

In order to have a better understanding of the contribution of each scale to the total Reynolds stress we will
use the marginal contribution of energy (MCE) defined in \cite{Pozuelo_PRF_arxiv, Pozuelo_JFM_22} and applied to the streamwise velocity component.
For each wall-normal position, the MCE yields the percentage of $\overline{u^2}$ that is provided by the small scales up to a certain wavenumber. In Eq.~(\ref{eq:cumsum}), it can be observed that MCE is a function of $y$ and $k_z$. 
\begin{equation}
\mathrm{MCE}(y,k_{z,c}) = \int_{k_{z,c}}^{\infty} \phi_{uu} \mathrm{d} k_{z} \ \bigg/ \int_{0}^{\infty} \phi_{uu} \mathrm{d} k_z.
\label{eq:cumsum}
\end{equation}
The integral in the numerator accumulates the energy of the smallest scale with $k_z \rightarrow \infty$ up to the cut-off frequency $k_{z,c}=k_z$.
The integral at the bottom of the fraction spans all the scales and is equal to $\overline{u^2}$.
By representing the iso-contours of the MCE it is possible to see how relevant is the contribution of large and small scales at different wall-normal locations.
Parallel iso-contours for consecutive levels indicate that those scales have a similar energetic mechanism.
Non-parallel iso-contours are connected with a region where the scales have a different energetic contribution.
When the same iso-contour is compared for different simulations, if the curves are parallel it can be stated that a similar energetic behaviour of those scales is found, and if they are separated there is an indication that in one simulation the larger scales are more relevant than in the other one.
In our channel simulations, which share the Reynolds number and the same energy mechanisms, the differences in the iso-contours provide a visual representation of the regions most significantly affected by the different widths of the domain, both qualitatively and quantitatively.

\begin{figure}
\centering
\includegraphics[width=0.46\textwidth]{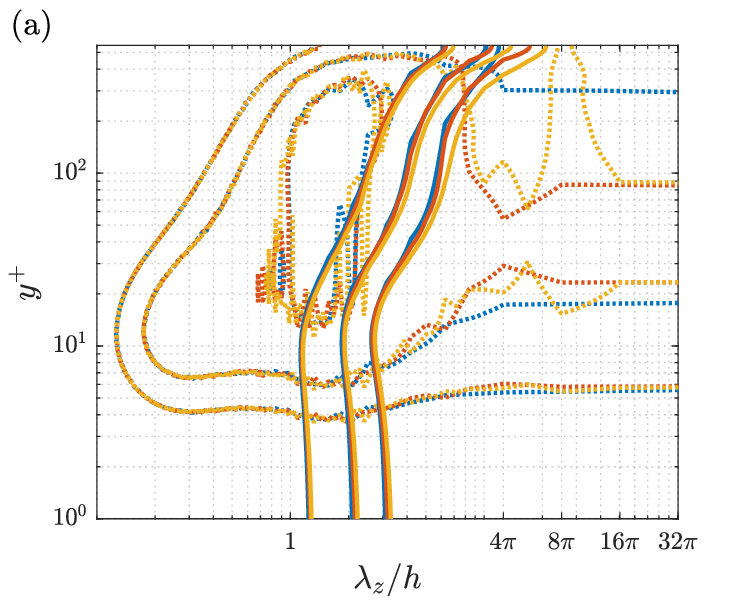}
\includegraphics[width=0.46\textwidth]{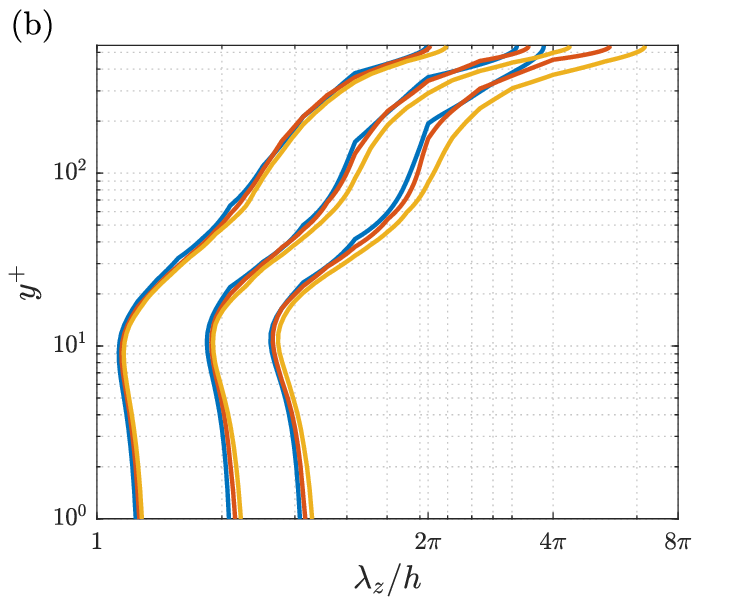}
\caption{(a) Power-spectral-density contours in dotted lines for levels $\phi_{uu}^+ = [0.05, 0.1, 0.2]$ and MCE contours in solid lines for levels $90\%$, $95\%$ and $97\%$ from left to right, shown for the channels C1, C2 and C3. (b) Detail of the MCE contours. Colors shown in Table~\ref{tab:param_sch}.}
\label{fig:PSD_MCE}
\end{figure}

In Fig.~\ref{fig:PSD_MCE} the MCE contours show that up to $90\%$ of the energy provided by the smaller scales remains unaltered by the width of these domains, with the widest streamwise scales of size $\lambda_z \approx 2\pi$ being at the center of the channel.
The effects of the width of the domain starts to be seen mostly in the outer region for the $95\%$ contours and for the $97\%$ contour the channel C1 already exhibits scales of size $\lambda_z/h = 4\pi$ at the center line, while channel C2 exhibits scales with $\lambda_z/h = 17$ and in channel C3 there are scales with $\lambda_z/h \approx 21$.
The viscous sublayer is slightly affected by the width of the domain, since the largest scales in the streamwise component can extend their influence down to the region close to the wall. 

\section{Conclusions} \label{sec:Conclusion}
The effects of the widest structures in turbulent channel flows have been assessed through turbulence statistics and spectral analysis of the fluctuations.
As expected, the smallest channel C1 is wide enough to properly reproduce the turbulence statistics, 
exhibiting very small differences with respect to the widest channel C3 which are of the order of the convergence error of our simulations.
The unclosed contours on the power-spectral density correspond to the region of low wavenumbers, or largest scales.
The power spectra for $k_z=0$ represent the variance of the average in the spanwise direction and is different from zero in the various cases. 
The wider the channel, the smaller is the energy contained in the largest modes, however, the power-spectral density becomes independent of the spanwise size of the domain.
This result is in agreement with what has been observed in turbulent boundary layers, and it could indicate the presence of spanwise-coherent hydrodynamic waves as mentioned in Abreu {\it et al.}~\cite{Abreu_JFM_2021}. 
The effect of the wider scales affects the flow all across the channel, and the widest resolved modes are those more affected by the domain size.


\section*{Acknowledgements}
RV acknowledges the financial support provided by the Swedish Research Council (VR).
The computations and data handling were enabled by resources provided by the National Academic Infrastructure for Supercomputing in Sweden (NAISS), partially funded by the Swedish Research Council.




\bibliography{jfm-instructions}


\end{document}